\documentclass[twocolumn,aps,prl,superscriptaddress,showpacs]{revtex4}
\usepackage{amssymb}
\usepackage{amsmath}
\usepackage{amsfonts}
\usepackage{graphicx}

\begin{document}

\title{Ultrafast mapping of optical polarization states onto spin coherence \\
of localized electrons in a semiconductor}

%\altaffiliation[Electronic mail:]{s.denega@rug.nl}
%Lines break automatically or can be forced with \\
\author{S.~Z.~Denega}
\author{M.~Sladkov}
\affiliation{Zernike Institute for Advanced Materials,
University of Groningen, NL-9747AG Groningen, The
Netherlands}
\author{D.~Reuter}
\author{A.~D.~Wieck}
\affiliation{Angewandte Festk\"{o}rperphysik, Ruhr-Universit\"{a}t
Bochum, D-44780 Bochum, Germany}
\author{T.~L.~C.~Jansen}
\author{C.~H.~van~der~Wal}
\affiliation{Zernike Institute for Advanced Materials,
University of Groningen, NL-9747AG Groningen, The
Netherlands}

%%%%%%%%%%%%%%%%%%%%%%%%%%% with "Nijenborgh 4"
%\affiliation{Zernike
%Institute for Advanced Materials,\\
%University of Groningen, Nijenborgh 4, NL-9747AG Groningen,
%The Netherlands}

%\date{April 15, 2010}
\date{\today}

\begin{abstract}
We experimentally demonstrate an ultrafast method for preparing spin states of donor-bound electrons in GaAs with single laser pulses.
Each polarization state of a preparation pulse has a direct mapping onto a spin state, with bijective correspondence between the Poincar\'{e}-sphere (for photon polarization) and Bloch-sphere (for spin) state representations. The preparation is governed by a stimulated Raman process and occurs orders of magnitude faster than the spontaneous emission and spin dephasing.
Similar dynamics governs our ultrafast optical Kerr detection of the spin coherence, thus getting access to spin state tomography. Experiments with double preparation pulses show an additive character for the preparation method. Utilization of these  phenomena is of value for quantum information schemes.
%for using this system in quantum information schemes.
%Thus, any spin superposition can be prepared.
\end{abstract}

\pacs{42.50.Ex, 42.50.Dv, 42.50.Ct, 71.55.Eq}
%42.50.Ct	Quantum description of interaction of light and matter; related experiments
%
%42.50.Dv	Quantum state engineering and measurements (see also 03.65.Ud Entanglement and quantum nonlocality, e.g., EPR paradox, Bells inequalities, GHZ states, etc.)
%
%42.50.Ex	Optical implementations of quantum information processing and transfer
%
%42.50.Gy	Effects of atomic coherence on propagation, absorption, and amplification of light; electromagnetically induced transparency and absorption
%
%42.50.Hz	Strong-field excitation of optical transitions in quantum systems; multiphoton processes; dynamic Stark shift (for multiphoton ionization and excitation of atoms and molecules, see 32.80.Rm, and 33.80.Rv, respectively)
%
%
%42.50.Md	Optical transient phenomena: quantum beats, photon echo, free-induction decay, dephasings and revivals, optical nutation, and self-induced transparency
%... ... ...	Dynamics of nonlinear optical systems; optical instabilities, optical chaos, and optical spatio-temporal dynamics, see 42.65.Sf
%... ... ...	Optical solitons; nonlinear guided waves, see 42.65.Tg

%71.55.-i	Impurity and defect levels
%
%71.55.Ak	Metals, semimetals, and alloys
%
%71.55.Cn	Elemental semiconductors
%
%71.55.Eq	III-V semiconductors

\maketitle

The ability to rapidly prepare and detect arbitrary spin states is of key interest in the fields of quantum information \cite{ladd2010} and spintronics \cite{Awschalom2009}.
This is currently explored with the electron spin in many different material systems such as atomic gases \cite{Duan2001,Chou2005,Chaneliere2005,muller2009}, quantum dots \cite{Koppens2006,Press2008,Greilich2009,Berezovsky2008,Schmidgall2010,Wu_prl_2011} and defect centers in diamond \cite{Fuchs2010,Togan2010,DeLange2010}. Donor-bound electrons in semiconductors ($D^{0}$ systems) are another interesting system \cite{Fu2005,Fu2006,Wang2007,Fu2008,Clark2009,Sladkov2010},
as they provide optically active centers with atom-like properties in solid state: $D^{0}$ systems combine a high level of homogeneity for ensembles (as for atomic vapors) with strong optical transitions and the ability to nano-fabricate and integrate very compact optoelectronic devices with semiconductor processing tools.

We report here a method for preparation and detection of arbitrary $D^{0}$ spin states with picosecond laser pulses, using GaAs with Si donors at very low concentration.
The polarization state of the pump pulse fully determines which spin state is prepared. Such preparation \cite{Kosaka2008} and tomographic detection \cite{Kosaka2009} was recently also reported for a GaAs quantum well system. However, there spin coherence was prepared for photo-electrons from single-photon transitions, such that the coherence was limited by electron-hole recombination ($\sim$300~ps). We developed an analogous method, but with two-photon Raman transitions between the two $D^0$ spin states, such that spin coherence is prepared for donor-bound electrons and thus not limited by electron-hole recombination.

% fig.1
%\newpage
\begin{figure}[b!]
%\begin{center};
% SERGII ORIGINAL FIG HAD WIDTH 80mm
\includegraphics[width=86mm]{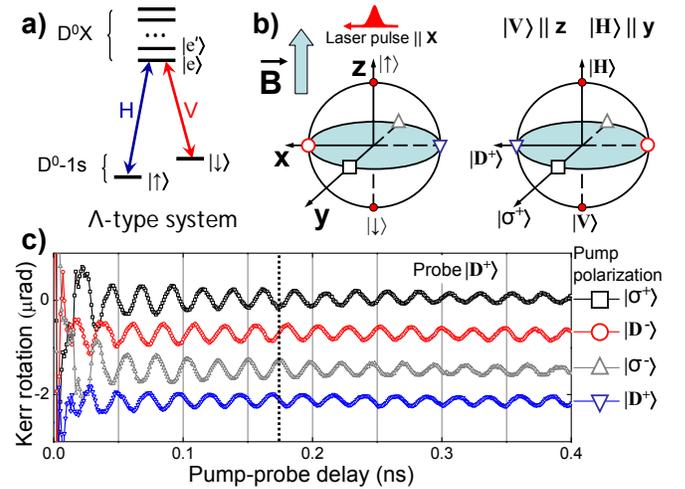}
%\end{center}
\caption{(Color online)
    (a)~Energy levels and optical transitions of the $D^{0}$-$D^{0}X$ system.
    (b)~Bloch sphere representation of electron spin states (left) and Poincar\'{e}-sphere representation of the photon polarization states (right). The field $\vec{B}$ causes spin precession about the $z$-axis. Laser pulses propagate along the $x$-axis and drive spin state preparation, and subsequent detection with a delayed pulse.
    (c)~Time-resolved Kerr rotation signals displaying the time evolution of four different initial states in the Bloch-sphere equatorial plane (offset for clarity). The phase of the oscillations in each trace corresponds to the precession phase of the spin state. The initial spin state is determined by the polarization of the preparation pulse,
    as labeled with matching symbols in (b) and (c).
    Comparing the oscillations at --for example-- the dashed line (at 175~ps delay) shows four different phase values.}
    \label{fig:fig1}
\end{figure}

Preparing pure spin-up and -down states by optical pumping was already demonstrated with the GaAs $D^0$ system \cite{Fu2006,Fu2008}. Preparation of arbitrary spin states was demonstrated using Coherent Population Trapping (CPT) \cite{Fu2005,Sladkov2010}, but the spectroscopic character of this method makes it difficult to control the phase of the quantum state. Moreover, in both cases the preparation takes a few cycles of the spontaneous emission time ($\sim$1~ns). Work with detuned picosecond laser pulses demonstrated all-optical spin manipulation \cite{Fu2008} and spin echo \cite{Clark2009} with a spin coherence time $T_2$ in excess of $20~\mu s$. Our work here adds picosecond preparation and detection to the available operations for $D^0$ spins, thus expanding a set with a ratio between coherence time and operation time in excess of $10^{7}$. Completing such a set of tools is essential for investigating quantum error correction schemes \cite{nielsen2000book}.

We explored the preparation of spin states with laser pulses that are resonant with transitions of the $D^0$ system to states with an additional electron-hole pair (donor-bound trion system $D^0X$). However, with fast pulses it is impossible to address transitions to the lowest $D^0 X$ level $|e\rangle$ only, since within the spectral width of a short laser pulse it has many energy levels [Fig.~\ref{fig:fig1}(a)]. The selection rules for these transitions are --despite extensive studies \cite{Karasyuk1994,Fu2005,Sladkov2010}-- not yet fully understood. The coupling to these higher levels can play a role, as well as weak coupling to other solid-state excitations \cite{Fu2008}. This situation is therefore difficult to assess theoretically. Experimentally, we find nevertheless remarkably simple and robust behavior for the preparation and detection. The results systematically show the behavior of an effective three-level system, and the associated quantum optical effect of CPT that was observed in $D^0$ studies with continuous wave (CW) lasers \cite{Fu2005,Sladkov2010}. While the higher $D^0X$ levels can be modeled with one effective $|e'\rangle$ level \cite{Wang2007}, we have at this stage no complete understanding of the fact that these levels have such little influence on the spin preparation and detection step.

%  This was motivated by observations that red detuned pulses require high power levels that induce rapid dephasing \cite{Fu2008}, and that resonant operation is in practice less sensitive to power fluctuations \cite{Wu_prl_2011}.
%We focus here on experimental investigation,
%and leave full theoretical description for future work.

We therefore describe the $D^0$ system as a typical three-level $\Lambda$-scheme \cite{Fu2008} for which a spin-up
and -down state ($\mid\uparrow\rangle$ and
$\mid\downarrow\rangle$) both have an optical transition to the same excited state $|e\rangle$ [Fig.~\ref{fig:fig1}(a)]. The degeneracy of $\mid\uparrow\rangle$ and $\mid\downarrow\rangle$ is lifted by an applied magnetic field $|\vec{B}|=7$~T along the $z$-axis.
The state $|e\rangle$ is then the lowest energy level of the $D^0 X$ complex, with two electrons in a singlet state and a hole with $m_{h}=-\frac12$ \cite{Clark2009}. We use optical propagation orthogonal to $\vec{B}$ (Voigt geometry). High-resolution spectroscopy for this case \cite{Sladkov2010} showed selection rules where the $\mid\uparrow\rangle$-$|e\rangle$ transition only couples to
H-polarized light (linear polarization orthogonal to $\vec{B}$), and the $\mid\downarrow\rangle$-$|e\rangle$ only to V-polarized light (parallel with $\vec{B}$). The polarization of light will further be denoted as quantum states $|H\rangle$, $|V\rangle$, and we also use the notation $|D^\pm\rangle=\frac{1}{\sqrt{2}}(|H\rangle\pm|V\rangle)$ and $|\sigma^\pm\rangle=\frac{1}{\sqrt{2}}(|H\rangle\pm i|V\rangle)$.

% fig.2
%\newpage
\begin{figure}[t!]
%\begin{center};
\includegraphics[width=80mm]{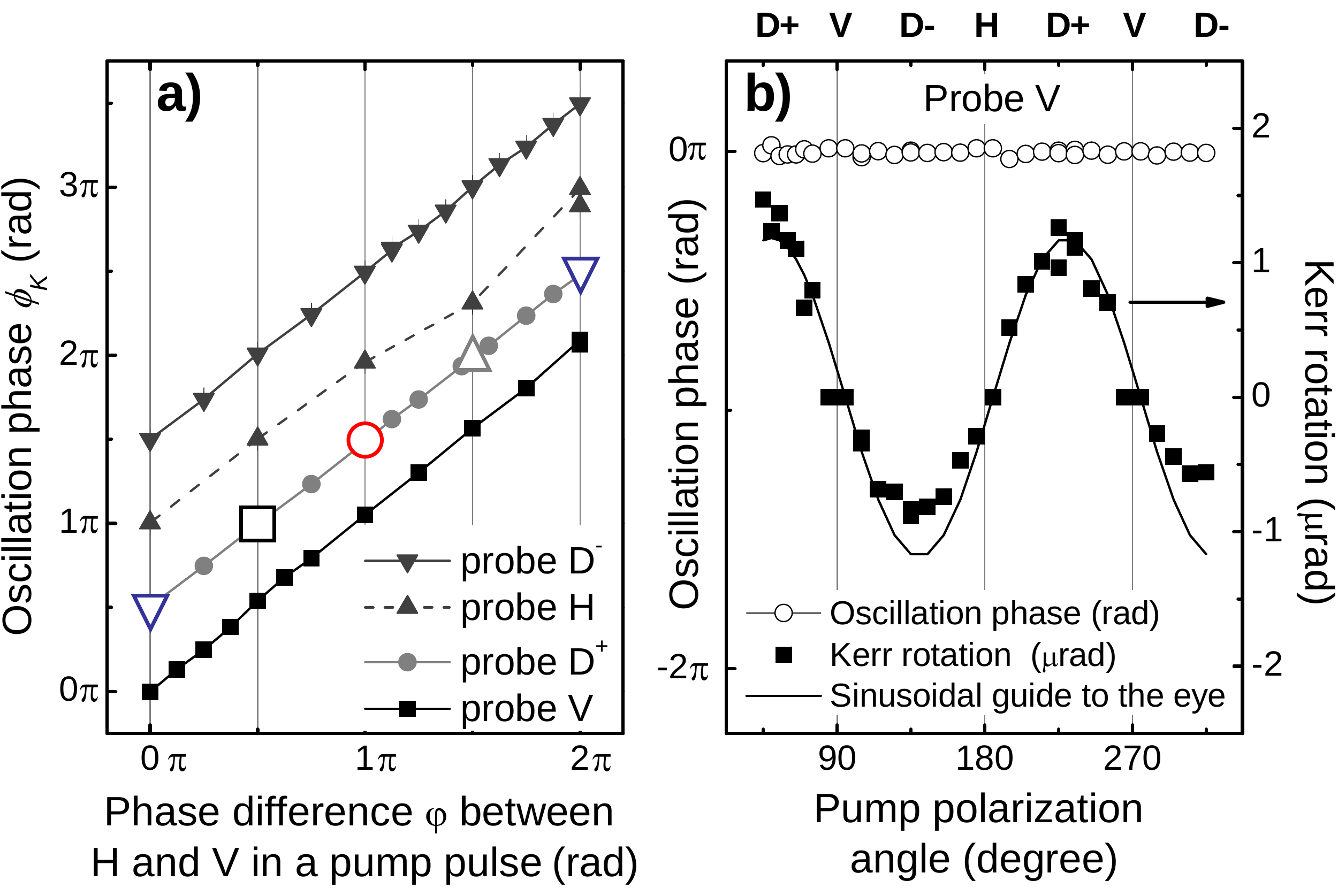}
%\end{center}
\caption{(Color online) (a)~The phase $\phi_K$ of oscillations in Kerr signals for different probe polarizations, as a function of the phase difference $\varphi$ between the $H$- and $V$-component in the preparation pulse. Empty symbols correspond to the data extracted from Fig.~\ref{fig:fig1}(c).
(b)~Amplitude (right axis) and phase (left axis) for the oscillating Kerr signal as a function of polarization angle (see also top axis) for a linearly polarized preparation pulse (with --unlike (a)-- using in the fits a negative amplitude rather than a phase change of $\pi$).}\label{fig:fig2}
\end{figure}

The mentioned CPT physics results from destructive quantum interference in the system's dynamics when driving the H- and V-transition at the same time. The system then gets trapped in a coherent superposition of the $\mid\uparrow\rangle$ and
$\mid\downarrow\rangle$ states only \cite{Fleischhauer2005,Fu2005,Sladkov2010}. This state is proportional to $\Omega_{V} \mid\uparrow\rangle - \Omega_{H} \mid\downarrow\rangle$,
where the complex optical Rabi frequencies $\Omega_{H}$ and $\Omega_{V}$ account for both the amplitude and the phase of two selectively applied driving fields. In the conventional CPT case (driving with a pair of weak CW lasers), the transition from an incoherent initial state into the CPT state takes a few times the spontaneous emission time \cite{Fleischhauer2005}.
For our present work, however, we consider the case where we drive both transitions with a single picosecond laser pulse which has an H- and V-component, and with a spectral width ($\Delta E=1.8$~meV) that is larger than the $D^0$ Zeeman splitting ($E_Z=0.16$~meV). The polarization state of the pulse
$\alpha|H\rangle  +  e^{i\varphi} \beta |V\rangle$
thus determines how it interacts with the three-level system, and we find behavior that is consistent with the mapping
\begin{equation}
\alpha|H\rangle  +  e^{i\varphi} \beta |V\rangle
\;\;\;\;
\rightarrow
\;\;\;\;
\beta \mid \uparrow\rangle  -  e^{-i\varphi} \alpha \mid \downarrow\rangle ,
\label{eq:eqMap}
\end{equation}
where $\alpha$ and $\beta$ are positive real-valued probability amplitudes, and $\varphi$ is a phase for the quantum superpositions. The swap of $\alpha$ and $\beta$ and sign change for the phase directly reflect the underlying CPT physics. We thus operate a unique mapping from polarization state onto spin state with direct correspondence between the Poincar\'{e}-sphere (for polarization) and Bloch-sphere (for spin) state representations [Fig.~\ref{fig:fig1}(b)].

Typical results are presented in Fig.~\ref{fig:fig1}(c). We used a stroboscopic pump-probe method with Kerr detection (further discussed below) for time-resolved preparation and detection of $D^0$ spin coherence. Figure~\ref{fig:fig1}(c) shows oscillating Kerr signals that directly reflect the precession of prepared spin states about the field $\vec{B}$. Notably, the phase of the precession is determined by the polarization of the pump pulse, and can be traced back to an initial state that shows a unique mapping between polarization and spin states.

% fig.3
%\newpage
\begin{figure}[t!]
%\begin{center};
\includegraphics[width=80mm]{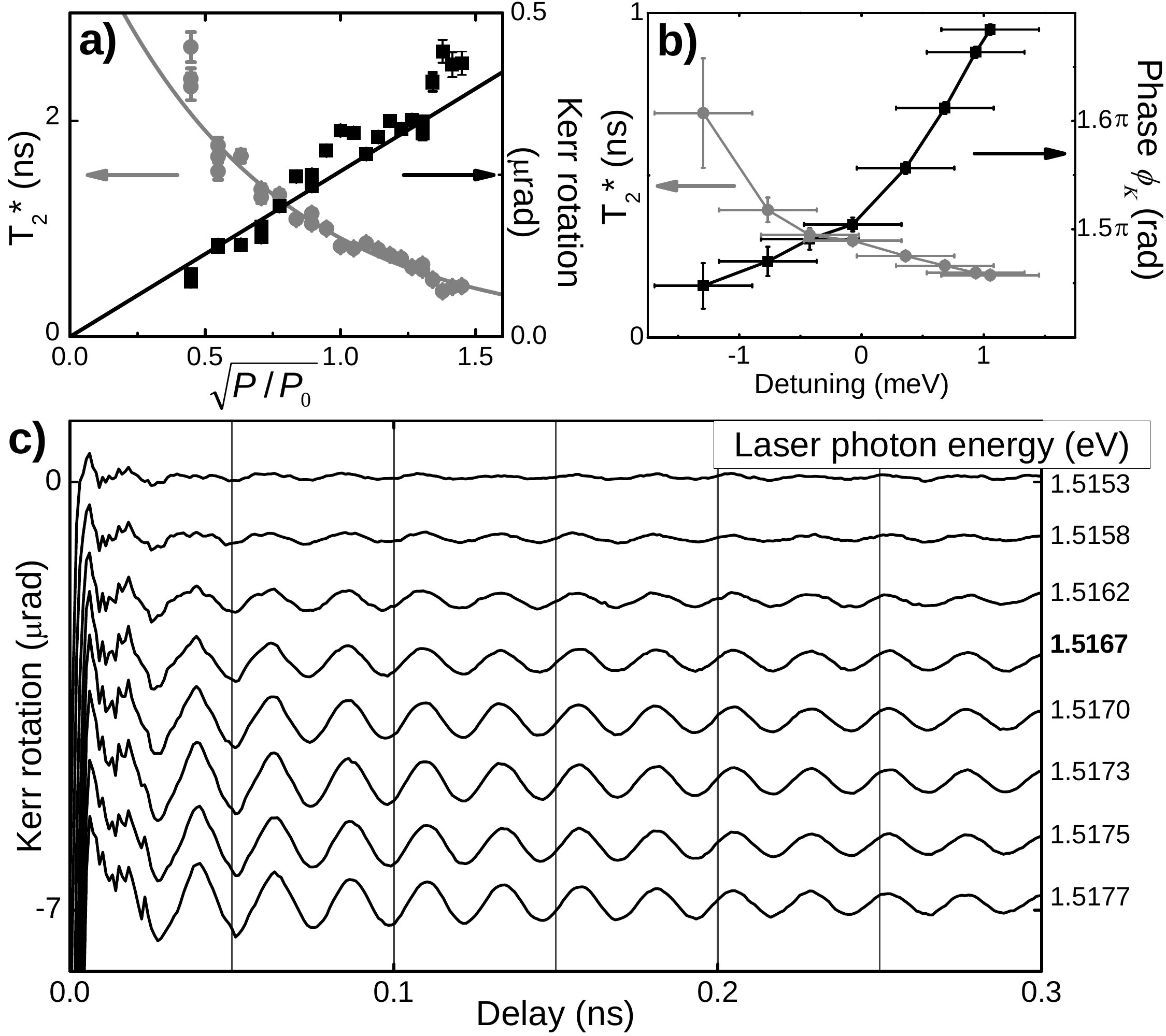}
%\end{center}
\caption{(a)~Dependence of the spin dephasing time $T_2^*$ (left axis, solid line is guide to the eye) and Kerr oscillation amplitude (right axis, solid line is linear fit) on preparation power $P$, where $P_{0}$=1~mW. (b)~Spin dephasing time $T_2^*$ and the Kerr oscillation phase $\phi_K$ as a function of laser detuning. (c)~Kerr signals for different excitation photon energies (tuned simultaneously for preparation and probe pulses). Data is offset for clarity. Polarizations of pump and probe beams are $|D^-\rangle$ and $|D^+\rangle$ respectively.}\label{fig:fig3}
\end{figure}

We used an epitaxial GaAs film grown along [001] of $10 \; {\rm \mu m}$ thickness with Si doping at $n_{\rm Si} \approx 3\times10^{13}~{\rm cm}^{-3}$ (same material as our CPT results \cite{Sladkov2010} but with the epitaxial lift-off process omitted). Measurements were performed at 4.2~K in an optical cryostat with superconducting magnet, and with pump and probe laser beams at normal incidence to the sample plane. The pump and probe beams were focused into a spot of 120~$\rm{\mu m}$ diameter.
The laser photon energy was $E=$1.5167~eV (unless stated otherwise) and tuned to resonance with transitions from $\mid\uparrow\rangle$ and $\mid\downarrow\rangle$ to $|e\rangle$. This photon energy is sufficiently low for avoiding the generation of free excitons. The spectral width of the laser pulses was reduced to 1.8~meV by filtering with a tunable liquid-crystal Fabry Perot (LCFP). All measurements were performed in the regime $T_{2}^{*}<T_{rep}<T_{1}$, where $T_{2}^{*}\approx$2~ns is the ensemble spin dephasing time, $T_{rep}\approx$12~ns is the laser repetition time, and $T_{1}\approx$1~ms is the $D^0$ spin relaxation time (value taken from Ref.~\cite{Fu2006}).

For determining the $D^0$ spin state we measure the Kerr rotation of a reflected probe pulse, while scanning a controllable delay $t$ between preparation and probe pulses. The polarization rotation is measured with a polarization bridge. Widespread application of this technique showed that it measures spin orientation along the probe beam when applied to a continuum of transitions in a semiconductor \cite{Kikkawa1997}. However, we apply it to the discrete set of transitions of the $D^{0}$-$D^{0}X$ system, and find that we can measure the $D^0$ spin state along any desired basis by choosing the probe polarization. This gives access to spin state tomography, analogous to the method that was developed for GaAs quantum wells \cite{Kosaka2009}.

Figure~\ref{fig:fig2}(a) presents results from a systematic study of phase values $\phi_K$ that appear in the oscillations of the measured Kerr signals, obtained with four different linear polarizations of the probe.
The phase values $\phi_K$ were obtained by fitting the signals to a function with factor
$\cos(\omega_L t + \phi_K)$ to represent the oscillations at the Larmor frequency $\omega_L$, and with a mono-exponentially decaying envelope. The $\phi_K$ values are studied as a function of a continues range of phase values $\varphi$ applied in preparation pulses with polarization states in the form
$\frac{1}{\sqrt{2}}(|H\rangle  +  e^{i\varphi} |V\rangle)$
(these are all states on the Poincar\'{e}-sphere equator, and the plot includes the results of Fig.~\ref{fig:fig1}(b,c) with matching symbols, for states that were prepared with phase values $\varphi=0$, $\pi/2$, $\pi$ and $3\pi/2$.
The four straight lines evidence that our method can prepare
any state on the equator in a manner that is consistent with the mapping as in Eq.~\ref{eq:eqMap}. In addition, the phase offset between the four straight lines confirms that the probe polarizations $|V\rangle$, $|D^+\rangle$, $|H\rangle$, $|D^-\rangle$ measure spin orientation along the $-x$, $-y$, $+x$, $+y$ direction, respectively. We focussed on measuring the equator states, as these carry a clear signature of spin precession for which we can analyze the phase after careful calibration of the point of zero pump-probe delay. Our results indicate that spin orientation along the $z$-axis can be measured with $|\sigma^\pm\rangle$ probe polarization, but this was more difficult to measure given the long $T_{1}$ time for our system.

For the preparation, we could confirm the mapping of Eq.~\ref{eq:eqMap} with states below or above the equator, with a continuous range of polarization states
$\alpha|H\rangle  + \beta |V\rangle$, with varying real $\alpha$ and $\beta$ and fixed $\varphi=0$. This shows, for example, a sinusoidal modulation of the amplitude of Kerr oscillations at a fixed phase for the signal of spin-orientation along the $-x$-axis [Fig.~\ref{fig:fig2}(b)].
This study thus confirms that our method shows bijective mapping between polarization states and spin states, both for preparation and detection.

Since the excitation occurs at energies which are below the GaAs band gap (and resonant with $D^0$ transitions), the presented Kerr signals are purely due to $D^0$ spin coherence (in the part of the signal beyond a delay of $\sim$25~ps). This is confirmed by the fact that the signals live longer than the various electron-hole recombination processes that can occur in this system. In addition, we find in all traces a $g$-factor of $|g|$=0.423$\pm$0.002, which is consistent with the $D^0$ case at 7~T. This provides evidence that both the preparation and detection must be due to a stimulated Raman process that occurs during the pulses.

We studied the robust character of our method by varying the intensities and detuning of the laser pulses. The amplitude of Kerr signals (directly representing spin coherence) in Fig.~\ref{fig:fig3}(a) appears to be proportional to the optical Rabi frequency
of the excitation (note that the scale represents actual measured Kerr rotation angle). It remains almost linear until 10$\cdot$$P_0$ (high-$P$ data not shown). This behavior is consistent with the presence of chirped modulation in the laser pulses \cite{Schmidgall2010,Wu_prl_2011} (which appears in our setup after spectral filtering of the laser with the LCFP).
The increase of the excitation power also reduces the spin dephasing time. Otherwise, our preparation and detection scheme are robust against a significant change in laser power.

% fig.4
%\newpage
\begin{figure}[t!]
%\begin{center};
\includegraphics[width=80mm]{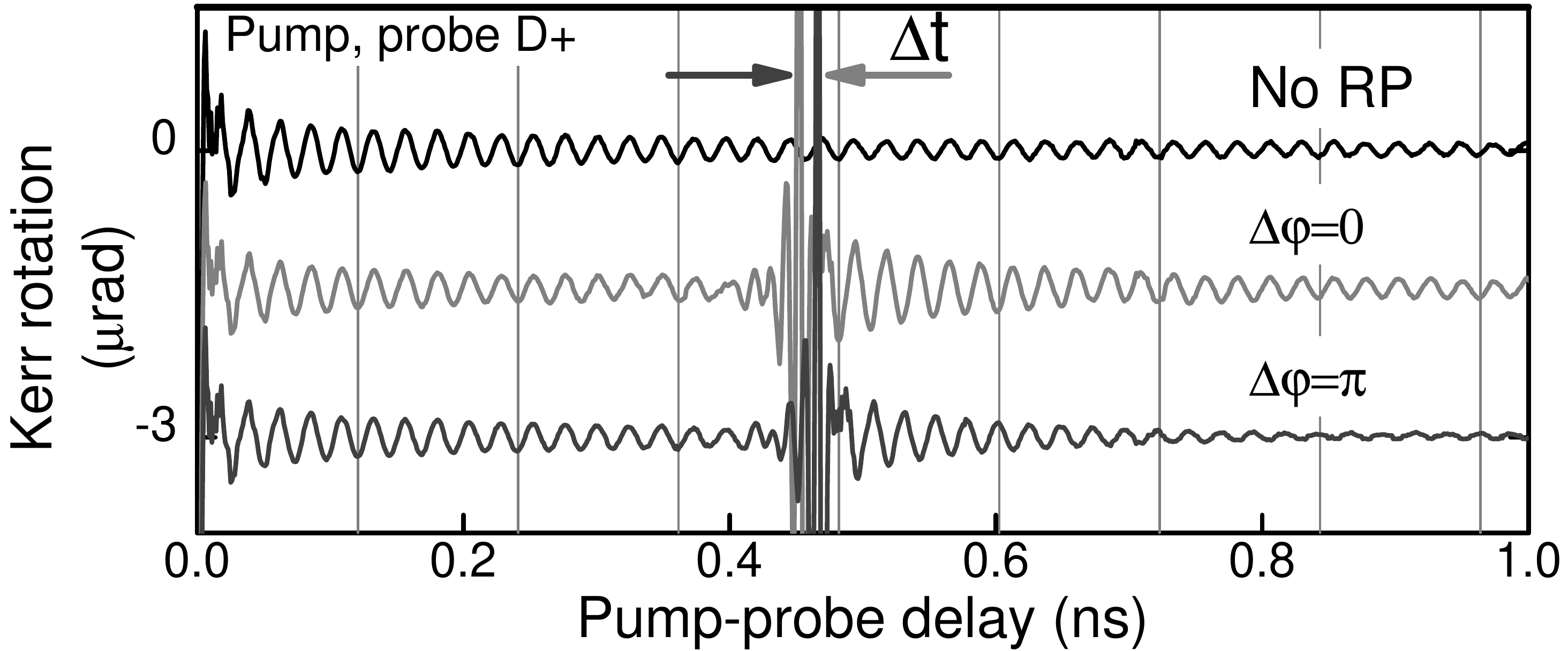}
%\end{center}
\caption{Kerr oscillation traces for various delays between initial and repeated preparation (RP) pulses (data offset for clarity).}\label{fig:fig4}
\end{figure}

By increasing the photon energy for the laser pulses from exact resonance we test how robust our spin preparation and detection method is with respect to an increased role for the higher levels $|e'\rangle$ of the $D^0X$ complex [Fig.~\ref{fig:fig1}(a)].
Increasing the photon energy results in stronger interaction with these levels, and Fig.~\ref{fig:fig3}(b) shows that a 1~meV blue detuning gives a 0.18~$\pi$ phase shift to the observed spin precession. At the same time, the efficiency of state preparation goes up (higher signal amplitude) while the spin dephasing time goes down [Fig.~\ref{fig:fig3}(b,c)].
This can be explained as follows: More levels involved in the Raman process gives a higher efficiency for the preparation process, but different polarization selection rules for the higher levels $|e'\rangle$ \cite{Sladkov2010} also cause a phase shift for the state that is prepared. The method therefore functions best and is most robust for pulses that are resonant or slightly red detuned from exact resonance with transitions to the lowest $D^0X$ level $|e\rangle$.

Further insight into the robust character of our preparation method can be obtained from applying a preparation pulse to an ensemble which already contains spin coherence \cite{Greilich2009}. To this end, we performed experiments with two identical preparation pulses in sequence, where the repeated pulse (RP) is reaching the system after a controllable delay with respect to the initial pulse. Figure~\ref{fig:fig4} shows the evolution of the spin state with and without applying the RP,
and for two RP delay values that differ by $\Delta t$. The RP can either enhance [Fig.~\ref{fig:fig4} middle trace] or suppress [bottom trace] the original spin signal [top trace], and this depends on the delay between the initial pulse and RP. In particular, the enhancement (suppression) occurs when the RP prepares a state that is in phase (in counter phase) with the ensemble's ongoing free precession, as labeled by $\Delta\varphi=0$ ($\Delta\varphi=\pi$).

Notably, the experimental data shows that the RP at our power levels does neither result in pure spin manipulation \cite{Fu2008,Greilich2009}, nor in pure preparation without memory effects. Instead, the RP results in a state that is the vector sum of the Bloch-sphere state vectors for the freely precessing state and the RP-induced state.
Since the RP comes when the original state is already partially dephased, the resulting state does not show full cancelation in the case of an RP that adds a state in counter phase.
The additive character of precessing and RP-pulse induced states points to the option to use $D^0$ ensembles for storing the algebraic sum of two or more quantum states.

In conclusion, we presented a picosecond spin preparation and detection method for $D^0$ electron spins, which gives a ratio of $10^7$ between the $T_2$ coherence time and operation time. Our preparation and detection method show a simple and robust bijective mapping between photon polarization states and $D^0$ spin states. While we have a strong indication that the mentioned CPT physics plays a role for the preparation step, we have at this stage less insight in the mechanism that underlies the detection step and can not provide a theoretical framework. It is also not yet clear to what degree our preparation method induces an incoherent population into the level $|e\rangle$ and higher levels. This would mean that we prepare --with respect to the full three-level system-- a partially coherent state in the
donor-bound electron systems. However, such states can still serve as a basis for quantum information operations
\cite{Gershenfeld1997}. An exchange of quantum optical signals between spin ensembles in different micronscale volumes could provide a path to creating entanglement between the state of different spin ensembles, but it remains to be explored how this can be worked out for ensembles that are partially incoherent.

We thank the Dutch NWO and NanoNed, and the German programs DFG-SFB 491, DFG-SPP 1285 and BMBF nanoQUIT for financial support.
%Netherlands Organization for Scientific Research (NWO)
%and NanoNed Network

% \bibliography{D:/docs/lib/library}

%\begin{thebibliography}{99}
%
%\bibitem{awschalom2009phys} D. D. Awschalom and N. Samarth, Physics \textbf{2}, 50 (2009).
%%, and references therein.
%
%\end{thebibliography}

\end{document}